\documentclass[10pt,A4paper,conference]{IEEEtran}

\usepackage{amsfonts}
\usepackage{amsmath}
\usepackage{amssymb}

\begin{document}
\newtheorem{definition}{Definition}
\newtheorem{theorem}[definition]{Theorem}
\newtheorem{lemma}[definition]{Lemma}

\def\im{\mathop{\mathrm{im}}\nolimits}
\def\tr{\mathop{\mathrm{tr}}\nolimits}
\def\det{\mathop{\mathrm{det}}\nolimits}

\title{
Quantum Information Theoretical Analysis of Various Constructions for
Quantum Secret Sharing
}

\author{\authorblockN{Karin Rietjens}
\authorblockA{Dep. of Math. and Comp. Science \\
Eindhoven University of Technology\\
The Netherlands \\
Email: k.p.t.rietjens@tue.nl}
\and
\authorblockN{Berry Schoenmakers}
\authorblockA{Dep. of Math. and Comp. Science \\
Eindhoven University of Technology\\
The Netherlands\\
Email: berry@win.tue.nl} \and
\authorblockN{Pim Tuyls}
\authorblockA{Information Security Systems\\
Philips Research Eindhoven\\
The Netherlands\\
Email: pim.tuyls@philips.com}}

\maketitle

\begin{abstract}
Recently, an information theoretical model for Quantum Secret Sharing (QSS) schemes was introduced. By using this model, we prove that pure state Quantum Threshold Schemes (QTS) can be constructed from quantum MDS codes and vice versa. In particular, we consider stabilizer codes and give a constructive proof of their relation with QTS. Furthermore, we reformulate the Monotone Span Program (MSP) construction according to the information theoretical model and check the recoverability and secrecy requirement. Finally, we consider QSS schemes which are based on quantum teleportation.
\end{abstract}

\section{Introduction}
QSS schemes are used to share a quantum secret among a set of players such that only specific groups of players are able to reconstruct the secret (\emph{authorized sets}), while all other groups have no information about the secret at all (\emph{unauthorized sets}). The collection of unauthorized sets is called the \emph{adversary structure}, which has the property that every subset of an unauthorized set is also unauthorized.

In \cite{ImaiQuadeNascimentoTuylsWinter03}, an information theoretical model for a QSS scheme was defined. This model is used throughout the rest of this paper and is repeated here. Suppose one wants to share a secret $S$ which is an element of a $q$-dimensional Hilbert space $\mathcal{H}_{S}$, where $q$ usually is a prime power. The elements $\{ |0\rangle, |1\rangle, \ldots, |q-1\rangle \}$ form an orthonormal basis for $\mathcal{H}_{S}$ and we usually describe the state of the secret by its orthonormal decomposition $\rho_{S} = \sum_{i \in \mathbb{F}_{q}} \alpha_{i} |i\rangle \langle i|$. The reference system that purifies the state of $S$ is denoted by $R$ with corresponding Hilbert space $\mathcal{H}_{R}$. Finally, the secret is shared among a set of players $P = \{P_{1}, \ldots, P_{n} \}$ and the Hilbert space corresponding to a set $B \subseteq P$ is denoted by $\mathcal{H}_{B}$. The density matrix $\rho_{B}$ then describes the state of system $B$.

The model is defined as follows. We denote the mutual information between systems $R$ and $A$ by $\textsf{\emph{I}}(R:A) = \textsf{\emph{S}}(R) + \textsf{\emph{S}}(A) - \textsf{\emph{S}}(RA)$, where $\textsf{\emph{S}}(A)$ is the Von Neumann entropy of the state $\rho_{A}$ of system $A$.
\begin{definition}
\label{definition:quantum secret sharing protocol}
A \emph{QSS} scheme realizing an adversary structure $\mathcal{A}$ is described by a quantum operator which generates quantum shares from a quantum secret $S$ and distributes these among the players such that:
\begin{enumerate}
\item
\emph{recoverability requirement:}\\
for all $A \notin \mathcal{A}$ we have that $\textsf{\emph{I}}(R:A) = \textsf{\emph{I}}(R:S)$;
\item
\emph{secrecy requirement:}\\
for all $B \in \mathcal{A}$ we have that $\textsf{\emph{I}}(R:B) = 0$.
\end{enumerate}
\end{definition}
A scheme that satisfies these conditions is called a \emph{perfect} scheme. In a \emph{non-perfect} scheme, some sets have some information about the secret, but not enough to recover it, i.e.\ $0 \neq \textsf{\emph{I}}(R:B) < \textsf{\emph{I}}(R:S)$ for some unauthorized set $B$.\\

In this paper, we investigate Quantum Threshold Schemes (QTS) and their relation with Quantum Error Correcting Codes (QECC). In \cite{CleveGottesmanLo99}, it was shown that a $((t, 2t-1))$ QTS can be constructed from a $[[2t-1,1,t]]_{k}$ quantum code. Here, we give an information theoretical proof of this relation and also prove the reverse statement. In particular, we consider stabilizer codes and constructively show how these codes can be used for secret sharing. It is possible to compute the reduced density matrix of a subset of shares, by only making use of the properties of the stabilizer.

Furthermore, we reformulate the Monotone Span Program (MSP) construction \cite{Smith00} for a general adversary structure according to Definition \ref{definition:quantum secret sharing protocol}. By directly computing the reduced density matrix of a set of shares, we verify that the recoverability and secrecy requirement are satisfied.

Finally, the construction of a non-perfect $((n,n))$ QTS using teleportation, as was proposed in \cite{Bandyopadhyay00}, is reformulated in terms of the information theoretical model. We show that authorized sets satisfy the recoverability requirement, but unauthorized sets have some, but not enough, information about the secret.

\section{Pure and Mixed state QSS schemes}
In a \emph{pure state} scheme, the encoding of a pure state of the secret is a pure state, while with a \emph{mixed state} scheme the encoding of a pure state is sometimes a mixed state. In general, a QSS scheme is mixed, but it can be described as a pure scheme with one share discarded \cite{Gottesman99}. Therefore, it actually suffices to only consider pure state schemes, which have the following useful property. Part of it was previously considered in \cite{ImaiQuadeNascimentoTuylsWinter03}, but a full proof is given here.
\begin{theorem}
\label{theorem:purestate=>complementsauthorizedareunauthorized}
In a pure state QSS scheme, the recoverability requirement and the secrecy requirement are equivalent.
\end{theorem}
\begin{proof} Suppose $P$ is the set of all players and let $A, B \subseteq P$ such that $B = P \setminus A$. Using the Araki-Lieb inequality and the fact that the systems $RS$ and $RAB$ are in a pure state, we have
\begin{equation}
\textsf{\emph{I}}(R:S) - \textsf{\emph{I}}(R:A) = \textsf{\emph{I}}(R:B)
\end{equation}
and the theorem follows immediately.
\end{proof}
Note that this also implies that in a pure state scheme, the authorized sets are precisely the complements of the unauthorized sets and vice versa. Moreover, this implies that a pure state $((t,n))$ QTS satisfies $n = 2t-1$.

\section{QSS with Quantum MDS codes}
In the classical case, a linear $(t,n)$ threshold scheme over $\mathbb{F}_{q}$ can be constructed from an $[n+1,t,n+2-t]_{q}$ MDS code and vice versa \cite{McElieceSarwate81}. We show that in the quantum case, a $[[2t-1,1,t]]_{q}$ quantum MDS code can be used to construct a $((t,2t-1))$ QTS and vice versa. As a special case, binary stabilizer codes are considered and the recoverability requirement is checked by directly computing the entropies of the reduced density matrices of a subset of shares.\\

A quantum code can correct for erasures on a subsystem of the system of the codewords means that the operator that induces the erasures is perfectly reversible. The quantum data processing inequality \cite{SchumacherNielsen96} gives a necessary and sufficient condition for a quantum operator to be perfectly reversible.

A different condition for quantum erasure correcting is given by Theorem \ref{theorem:secrecyrequirementforQECC}. Cerf et al.\ \cite{CerfCleve97} previously proved the necessity of this condition. First we need the following lemma, which is given without proof.
\begin{lemma}
\label{lemma:subsystempure}
Let $A$ and $B$ be two quantum systems. If $A$ or $B$ is in a pure state, then the composite system $AB$ is in a product state.
\end{lemma}
\begin{theorem}
\label{theorem:secrecyrequirementforQECC}
Let $Q$ be a quantum system and let $R$ be its reference system, such that $RQ$ is in a pure state. Erasures can be corrected on some subsystem $Q_{e}$ of $Q$ if and only if $\textsf{\emph{I}}(R:Q_{e})=0$.
\end{theorem}
\begin{proof} Let $Q = Q_{u} Q_{e}$ and suppose we can correct for erasures on $Q_{e}$. This means that every quantum operator acting on $Q_{e}$ and leaving $Q_{u}$ invariant is perfectly reversible. Let $\mathcal{E}$ be a quantum operator that converts the system $Q_{e}$ into an arbitrary pure state and let $\rho_{R'Q'_{u}Q'_{e}}$ be the system $\rho_{RQ_{u}Q_{e}}$ after applying $I \otimes I \otimes \mathcal{E}$. Then $\rho_{R'Q'_{u}Q'_{e}}$ is in the product state $\rho_{R'Q'_{u}} \otimes \rho_{Q'_{e}}$ (Lemma \ref{lemma:subsystempure}) and $\textsf{\emph{S}}(R'Q'_{u}Q'_{e}) = \textsf{\emph{S}} (R'Q'_{u}) + \textsf{\emph{S}}(Q'_{e}) = \textsf{\emph{S}} (RQ_{u}) + \textsf{\emph{S}}(Q'_{e})$. Analogously, we have that $\textsf{\emph{S}}(Q_{u}Q'_{e}) = \textsf{\emph{S}} (Q_{u}) + \textsf{\emph{S}}(Q'_{e})$. Furthermore, because of the quantum data processing inequality we have $\textsf{\emph{S}}(Q) = \textsf{\emph{S}}(Q') - \textsf{\emph{S}}(R'Q')$ and therefore
\begin{eqnarray}
0 & = & \textsf{\emph{S}}(Q) - \textsf{\emph{S}}(Q') + \textsf{\emph{S}}(R'Q') \nonumber \\
& = & \textsf{\emph{S}}(R) - \textsf{\emph{S}}(Q_{u}Q'_{e}) + \textsf{\emph{S}}(RQ_{u}Q'_{e}) \nonumber \\
& = & \textsf{\emph{S}}(R) - \textsf{\emph{S}}(Q_{u}) - \textsf{\emph{S}}(Q'_{e}) + \textsf{\emph{S}}(RQ_{u}) + \textsf{\emph{S}}(Q'_{e}) \nonumber \\
& = & \textsf{\emph{S}}(R) - \textsf{\emph{S}}(RQ_{e}) + \textsf{\emph{S}}(Q_{e}) = \textsf{\emph{I}}(R:Q_{e}),
\end{eqnarray}
which completes the first part of the proof.

On the other hand, suppose $\textsf{\emph{I}}(R:Q_{e}) = 0$ for some subsystem $Q_{e}$ of $Q$. Let $\mathcal{E}$ be a quantum operator acting on $Q_{e}$. Then $\mathcal{E}$ has a representation as a unitary evolution on a larger system,
say $Q_{e}E$, where $E$ is initially in a pure state. Let $R'Q'E'$ be the system $RQE$ after this unitary evolution on $Q_{e}E$ and leaving $RQ_{u}$ invariant. Then because of the conservation rule for mutual information (see for example \cite{NielsenChuang00}) we have that
\begin{equation}
\textsf{\emph{I}}(R':Q'_{e}E') = \textsf{\emph{I}}(R:Q_{e}E).
\end{equation}
Since $E$ was initially in a pure state, we have (using Lemma \ref{lemma:subsystempure})
\begin{eqnarray}
\textsf{\emph{I}}(R:Q_{e}E) & = & \textsf{\emph{S}}(R) + \textsf{\emph{S}}(Q_{e}E) - \textsf{\emph{S}}(RQ_{e}E) \nonumber \\
& = & \textsf{\emph{S}}(R) + \textsf{\emph{S}}(Q_{e}) + \textsf{\emph{S}}(E) - \textsf{\emph{S}}(RQ_{e}) - \textsf{\emph{S}}(E) \nonumber \\
& = & \textsf{\emph{I}}(R:Q_{e}).
\end{eqnarray}
Furthermore, because of the strong subadditivity property for system $R'Q'_{e}E'$ we have
\begin{equation}
\textsf{\emph{I}}(R':Q'_{e}E') - \textsf{\emph{I}}(R':E') \ge 0,
\end{equation}
which implies that
\begin{equation}
0 \le \textsf{\emph{I}}(R':E') \le \textsf{\emph{I}}(R':Q'_{e}E') = \textsf{\emph{I}}(R:Q_{e}).
\end{equation}
Thus if $\textsf{\emph{I}}(R:Q_{e}) = 0$ then also $\textsf{\emph{I}}(R':E') = 0$, which is equivalent to $\textsf{\emph{S}}(Q) = \textsf{\emph{S}}(Q') - \textsf{\emph{S}}(R'Q')$ and therefore because of the quantum data processing inequality erasures can be corrected on $Q_{e}$. \end{proof}

Now we have the tools to prove the general relation between quantum MDS codes and QTS.
\begin{theorem}
\label{theorem:quantumMDSandthreshold}
A $((t, 2t-1))$ QTS, where the secret is an element of a $q$-dimensional Hilbert space, can be translated into a $[[2t-1,1,t]]_{q}$ quantum MDS code and vice versa.
\end{theorem}
\begin{proof} Consider a $((t,2t-1))$ QTS with system $S$ of the secret, reference system $R$ and system of the players $P$. Then the secrecy requirement states that for every set of at most $t-1$ players $B$, we have $\emph{\textsf{I}}(R:B)=0$. According to Theorem \ref{theorem:secrecyrequirementforQECC}, we have that erasures can be corrected on the shares of any set of $t-1$ players. Hence, all possible sets of shares in $P$ together form a $[[2t-1,1,t]]_{q}$ QECC.

On the other hand, consider a $[[2t-1,1,t]]_{q}$ quantum MDS code. We claim that each codeword can be the shares for a $((t, 2t-1))$ QTS. Indeed, if $Q$ is the composite system of the codewords and $R$ the reference system, then for every set $Q_{e}$ of at most $t-1$ of the $2t-1$ subsystems of $Q$ we have $\textsf{\emph{I}}(R:Q_{e}) = 0$ (Theorem \ref{theorem:secrecyrequirementforQECC}). Hence, the secrecy requirement is satisfied. Moreover, because of Theorem \ref{theorem:purestate=>complementsauthorizedareunauthorized} and the fact that a $((t,2t-1))$ QTS is a pure state scheme, we also have that the recoverability requirement is satisfied. \end{proof}

\subsection*{Stabilizer Codes}
We consider a $[[2t-1,1,t]]_{2}$ quantum stabilizer code with stabilizer $T$. We show that this code can be used to construct a $((t,2t-1))$ QTS and verify the recoverability requirement, which is sufficient because the scheme is pure. First, we present the following technical lemma.
\begin{lemma}
\label{lemma:stabilize->commute}
Let $W \in \mathcal{G}_{n}$, where $\mathcal{G}_{n}$ denotes the Pauli group on $n$ qubits, act on a composite quantum system $Q = Q_{1} \otimes Q_{2}$ and say $W=W_{1} \otimes W_{2}$, where $W_{i}$ acts on system $Q_{i}, i=1,2$. Suppose $|\psi\rangle$ is a state of system $Q$ that is stabilized by $W$. If $\rho_{2} = \tr_{1}(|\psi\rangle \langle \psi|)$, where $\tr_{1}$ is the trace over system $Q_{1}$, then $W_{2}$ and $\rho_{2}$ commute with each other.
\end{lemma}
\begin{proof} Let $A$ be an arbitrary quantum operator acting on the state space of system $Q_{2}$ and $\tr_{2}$ the trace over system $Q_{2}$. Then we have
\begin{eqnarray}
\tr_{2}(\rho_{2}A)
& = & \tr_{2} \Bigl( \tr_{1} (W |\psi\rangle \langle \psi| W^\dag) A \Bigr) \nonumber \\
& = & \tr_{2} \Bigl( W_{2} \rho_{2} W_{2}^\dag A \Bigr),
\end{eqnarray}
where we have used that the trace function is cyclic and the fact that $W_{1}^\dag W_{1} = I$ if $W_{1}$ is a tensor product of Pauli matrices. Since this holds for any $A$ acting on the state space of system $Q_{2}$ we have that $\rho_{2} = W_{2} \rho_{2} W_{2}^\dag$ which completes the proof.\end{proof}

Next let $T$ be generated by $\{ G_{1}, \ldots, G_{2t-2} \}$ and let $\overline{X}$ and $\overline{Z}$ be the logical Pauli $X$ and $Z$ operators on the logical basis $\{|0_{L}\rangle, |1_{L}\rangle\}$ for the stabilizer code (see \cite{NielsenChuang00}). Then $\{ G_{1}, \ldots, G_{2t-2}, \overline{X}, \overline{Z} \}$ forms a basis for the commutator $C(T)$ of $T$. Since an MDS code is pure \cite{Rains97}, we have that $T$ has minimum distance $t+1$ and $C(T)$ minimum distance $t$. This results in the following property, which we mention here without proof.
\begin{lemma}
\label{lemma:traceoutstillindependent}
If we restrict the generators of $C(T)$ to at most $t-1$ qubit positions, then the restricted generators of $C(T)$ remain independent.
\end{lemma}

We claim that the construction for the $((t,2t-1))$ threshold scheme is given by the following isometry.
\begin{definition}
The mapping $V_{t,2t-1} : \mathbb{C}^{2} \rightarrow (\mathbb{C}^2)^{\otimes 2t-1}$ is defined by
\begin{eqnarray}
V_{t,2t-1}(\alpha_{0}|0\rangle + \alpha_{1}|1\rangle) & = & \alpha_{0}|0_{L}\rangle + \alpha_{1}|1_{L}\rangle,
\end{eqnarray}
where $\alpha_{0}, \alpha_{1} \in \mathbb{C}$.
\end{definition}
So if the secret is in state $\rho_{S} = \alpha_{0} |0\rangle \langle0| + \alpha_{1} |1\rangle \langle 1|$, the state of the system of the shares $P$ is given by
\begin{equation}
\label{eq:stabilizerrhoPgeneralization}
\rho_{P} = V_{t,2t-1} \rho_{S} V_{t,2t-1}^\dag = \alpha_{0} |0_{L}\rangle \langle 0_{L} | + \alpha_{1} |1_{L} \rangle \langle 1_{L} |.
\end{equation}

The entropy of every possible subset of shares from $P$ is given by the following lemmas.
\begin{lemma}
\label{lemma:unauthorized}
Let $B \subset P$ with $|B| = t' \le t-1$. Then we have for the entropy of the state $\rho_{B}$ of system $B$
\begin{equation}
\textsf{\emph{S}} (B) = t' \log 2.
\end{equation}
\end{lemma}
\begin{proof}
Suppose $B$ is a set of $t-1$ qubits. Let $G'_{j}$ be the operator $G_{j}$ restricted to the qubit positions of $B$ for every $1 \le j \le 2t-2$. Because of Lemma \ref{lemma:stabilize->commute}, these operators $G'_{j}$ all commute with $\rho_{B}$. Moreover, because of Lemma \ref{lemma:traceoutstillindependent}, the operators $G'_{j}$ are still independent. Since $\rho_{B}$ is a $2^{t-1} \times 2^{t-1}$ density matrix that commutes with $2t-2$ independent elements in $\mathcal{G}_{t-1}$ we have that $\rho_{B} = 1/2^{t-1} I$. In general, for any set $B$ of at most $t-1$ shares, say $t'$, we have that $\rho_{B} = 1/2^{t'} I$.
\end{proof}

\begin{lemma}
\label{lemma:justauthorized}
Let $A \subset P$ with $|A| = t$. Then we have for the entropy of the state $\rho_{A}$ of system $A$
\begin{equation}
\label{eq:entropyt}
\textsf{\emph{S}} (A) = \textsf{\emph{S}}(S) + (t-1) \log 2.
\end{equation}
\end{lemma}
\begin{proof}
Consider a set $A$ of $t$ shares. Let $G'_{j}$, $\overline{X}'$ and $\overline{Z}'$ be the operators $G_{j}$, $\overline{X}$ and $\overline{Z}$ restricted to the qubit positions in $A$ respectively for every $1 \le j \le 2t-2$. Then these $2t$ operators are independent because of Lemma \ref{lemma:traceoutstillindependent}. Since $|0_{L}\rangle \langle 0_{L}|$ and $|1_{L}\rangle \langle 1_{L}|$ commute with $G'_{j}$ for every $j$ and also with $\overline{Z}$, we can write
\begin{eqnarray}
\rho_{A}^{0} \ = \ \tr_{A^{c}} (|0_{L}\rangle \langle 0_{L}|) & = & \frac{1}{2^t} I^{\otimes t} + \beta_{0} R; \\
\rho_{A}^{1} \ = \ \tr_{A^{c}} (|1_{L}\rangle \langle 1_{L}|) & = & \frac{1}{2^t} I^{\otimes t} + \beta_{1} R,
\end{eqnarray}
where $R \in \{I, X, Y, Z\}^{\otimes t}, R \neq I^{\otimes t}$ and $0 < |\beta_{0}|, |\beta_{1}| \leq 1/2^t$. The operator $R$ cannot commute with $\overline{X}'$, since then it would commute with $2t$ independent operators, which would imply that $R = I^{\otimes t}$. Therefore, since $R$ and $\overline{X}'$ are tensor products of Pauli matrices, $R$ anti-commutes with $\overline{X}'$. Hence, because $\overline{X'} \tr_{A^{c}} (  |0_{L}\rangle \langle 0_{L}|) \overline{X}'^\dag = \tr_{A^{c}} ( |1_{L}\rangle \langle 1_{L}| )$, we have that $\beta_{0} = -  \beta_{1}$.

Furthermore, $\rho_{A}^{0}$ has $2^{t-1}$ eigenvalues equal to $1/2^t + \beta_{0}$ and $2^{t-1}$ equal to $1/2^t - \beta_{0}$, because $R$ has $2^{t-1}$ eigenvalues equal to $+1$ and $2^{t-1}$ equal to $-1$. We also know that $\textsf{\emph{S}} (\rho_{A}^{0}) = \textsf{\emph{S}} ( \rho_{A^{c}}^{0} ) = (t-1) \log 2$, since $|0_{L}\rangle \langle 0_{L}|$ has zero entropy. Therefore, we have that $\beta_{0} = \pm 1/2^t$. Analogously for $\beta_{1}$.

Finally, by using the fact that $\alpha_{0} + \alpha_{1} =1$, we are able to verify that the entropy of $\rho_{A}$ is given by Eq. (\ref{eq:entropyt}), since $\rho_{A}$ has $2^{t-1}$ eigenvalues equal to $1/2^{t}(1+\alpha_{0}-\alpha_{1})$ and $2^{t-1}$ equal to $1/2^{t}(1-\alpha_{0}+\alpha_{1})$. \end{proof}

\begin{lemma}
Let $A \subseteq P$ with $|A| \ge t$. Then
\begin{equation}
\textsf{\emph{S}}(A) = \textsf{\emph{S}}(S) + (2t-1 - |A|) \log 2.
\end{equation}
\end{lemma}
\begin{proof}
Write $A = A_{t} \cup A'$, where $|A_{t}| = t$ and $|A'| = |A| - t \le t-1$. Let $B = P \setminus A$. Then by using Lemmas \ref{lemma:unauthorized} and \ref{lemma:justauthorized} we have
\begin{eqnarray*}
\textsf{\emph{S}}(A) & \ge & | \textsf{\emph{S}}(A_{t}) - \textsf{\emph{S}}(A') | \\
& = & \textsf{\emph{S}}(S) + (2t-1 - |A|) \log 2, \\
\textsf{\emph{S}}(A) & = & \textsf{\emph{S}}(RB) \\
& \le & \textsf{\emph{S}}(R) + \textsf{\emph{S}}(B) \\
& = & \textsf{\emph{S}}(S) + (2t-1 - |A|) \log 2,
\end{eqnarray*}
which completes the proof.
\end{proof}

Finally, we have the following.
\begin{theorem}
A $[[2t-1, 1,t]]$ binary stabilizer code can be used to share a secret according to a $((t,2t-1))$ QTS.
\end{theorem}
\begin{proof}
Let $A, B = P$, such that $B = P \setminus A$ and $|A| \ge t$. Then
\begin{eqnarray*}
\textsf{\emph{I}}(R:A) & = & \textsf{\emph{S}}(R) + \textsf{\emph{S}}(A) - \textsf{\emph{S}}(B) \\
& = & 2 \textsf{\emph{S}}(S) = \textsf{\emph{I}}(R:S).
\end{eqnarray*}
Hence, the recoverability requirement is satisfied for any authorized set of shares of Eq. (\ref{eq:stabilizerrhoPgeneralization}) and since the threshold scheme is pure, this completes the proof.
\end{proof}

\section{Monotone Span Program construction}
In \cite{Smith00} it was shown how (classical) MSP can be used to construct a QSS scheme for a general access structure. We show that the recoverability and secrecy requirement are fulfilled for this construction.\\

We only consider the pure state case. The recoverability and secrecy requirement for the mixed scheme follow immediately from the entropies for the pure scheme.

Let $\mathcal{A}$ be a self-dual adversary structure with corresponding MSP $(\mathbb{F}_{q},M,g)$ (see \cite{KarchmerWigderson93}), where $q$ is a prime power,  $M$ a $d \times e$ matrix over $\mathbb{F}_{q}$ with independent columns and $g$ a function that labels each row of $M$ with an element of $\{1, 2, \ldots, n\}$. Furthermore, by $\mathcal{H}$ we denote a $q$-dimensional Hilbert space and say the vectors that are labeled $\{ |\textbf{\emph{a}}\rangle \}_{\textbf{\emph{a}} \in \mathbb{F}_{q}^{n}}$ form an orthonormal basis for $\mathcal{H}^{\otimes n}$.

Consider the following isometry.
\begin{definition}
\label{definition:V_M}
The mapping $V_{M}: \mathcal{H}^{\otimes e} \rightarrow \mathcal{H}^{\otimes d}$ is defined by
\begin{equation}
V_{M} \Bigl( \sum_{i \in \mathbb{F}_{q}} \alpha_{i} | \psi^{i}_{1} \psi^{i}_{2} \ldots \psi^{i}_{e} \rangle \Bigr) = \sum_{i \in \mathbb{F}_{q}} \alpha_{i} \Biggl| M \left( \begin{array}{@{}c@{}} \psi^{i}_{1} \\ \psi^{i}_{2} \\ \vdots \\ \psi^{i}_{e} \end{array} \right) \Biggr\rangle,
\end{equation}
where $| \psi^{i}_{1} \psi^{i}_{2} \ldots \psi^{i}_{e} \rangle \in \mathcal{H}^{\otimes e}$ and $\alpha_{i} \in \mathbb{C}$ for every $i, 1 \le i \le q$.
\end{definition}
\ \\
We show that this mapping can be used to share a secret according to a QSS with adversary structure $\mathcal{A}$. Let the secret $S$ be an element of a $q$-dimensional Hilbert space $\mathcal{H}_{S}$ with orthonormal basis $\{|0\rangle, |1\rangle, \ldots, |q-1\rangle\}$. Again, $R$ denotes the reference system that purifies $S$ and $P$ denotes the system of the players. Let $I_{R}$ be the identity mapping on the system $R$. The encoding of the secret is then given by
\begin{equation}
\label{eq:MSPconstruction}
|RP\rangle = (I_{R} \otimes V_{M})(|RS\rangle \otimes |E\rangle),
\end{equation}
where
\begin{equation}
|E\rangle = \frac{1}{\sqrt{q^{e-1}}} \sum_{a \in \mathbb{F}_{q}^{e-1}} |\textbf{\emph{a}}\rangle
\end{equation}
and $\{ |\textbf{\emph{a}}\rangle \}_{\textbf{\emph{a}} \in \mathbb{F}_{q}^{e-1}}$ is an orthonormal basis for $\mathcal{H}_{E}$, the Hilbert space corresponding to system $E$.

This means that if the state of $S$ is described by the density matrix $\rho_{S}$, which has orthonormal decomposition
\begin{equation}
\rho_{S} = \sum_{i \in \mathbb{F}_{q}} \alpha_{i} |i\rangle \langle i|,
\end{equation}
then the state of the system of the shares $P$ together with the reference system $R$ is given by
\begin{equation}
|RP\rangle = \frac{1}{\sqrt{q^{e-1}}} \sum_{i \in \mathbb{F}_{q}} \sum_{\emph{\textbf{a}} \in \mathbb{F}_{q}^{e-1}} \sqrt{\alpha_{i}} \ |i\rangle \otimes \Bigl| M \left( \begin{array}{@{}c@{}} i \\ \emph{\textbf{a}} \end{array} \right) \Bigr\rangle.
\end{equation}
Finally, the dealer sends qudit $i$ to player $g(i)$ for $1 \le i \le d$.\\

Let $A$ be an authorized set and $B$ its unauthorized complement. To check the recoverability and secrecy requirement, we compute the entropy of system $A$ and $B$. By $M_{A}$ and $M_{B}$ we denote the rows of $M$ corresponding to $A$ and $B$ respectively, where $M_{A}$ has rank $l$ and $M_{B}$ rank $m$.

First, consider the following definition.
\begin{definition}
Let $B^{i}_{x}$ be the set of vectors $|i \ a_{1} \cdots a_{e-1}\rangle$ such that
\begin{equation}
M_{B} (i,a_{1}, \ldots, a_{e-1})^\top = \textbf{\emph{x}},
\end{equation}
where $i \in \mathbb{F}_{q},$ $\textbf{\emph{a}} = (a_{1}, \ldots, a_{e-1})^\top \in \mathbb{F}_{q}^{e-1}$ and $\textbf{\emph{x}} \in \im (M_{B})$. Then the vector $|\phi_{x}^{i}\rangle$ is defined by
\begin{equation}
\label{phi_p_i}
|\phi_{x}^{i}\rangle = \frac{1}{\sqrt{q^{e-m-1}}} \sum_{|i \emph{\textbf{a}}\rangle \in B_{x}^{i}} \Bigl| M_{A} \left( \begin{array}{@{}c@{}} i \\ \textbf{\emph{a}} \end{array} \right) \Bigr\rangle.
\end{equation}
\end{definition}
We claim that these vectors are the eigenvectors of the density matrix $\rho_{A}$ that describes the state of system $A$. To prove this, we need the next lemma.
\begin{lemma}
\label{lemma:oneelementincommon=>equalityB}
Consider two vectors $|\phi_{x}^{i}\rangle$ and $|\phi_{x'}^{i'}\rangle$ for certain $i, i' \in \mathbb{F}_{q}$ and $\emph{\textbf{x}},\emph{\textbf{x}}'\in \im(M_{B})$. Suppose there are vectors $|i \ \textbf{\emph{a}}\rangle \in  B_{x}^{i}$ and $|i' \ \textbf{\emph{a}}'\rangle \in B_{x'}^{i'}$ such that
\[
\Bigl| M_{A} \left( \begin{array}{@{}c@{}} i \\ \textbf{\emph{a}} \end{array} \right) \Bigr\rangle \ = \
\Bigl| M_{A} \left( \begin{array}{@{}c@{}} i' \\ \textbf{\emph{a}}' \end{array} \right) \Bigr\rangle .
\]
Then we have that $|\phi_{x}^{i}\rangle = |\phi_{x'}^{i'}\rangle$. If there are no such vectors, then $\langle \phi_{x}^{i} | \phi_{x'}^{i'}\rangle = 0$.
\end{lemma}
\begin{proof} It is sufficient to show that with the assumptions above, we have that for every $|i \ \textbf{\emph{b}}\rangle \in  B_{x}^{i}$, there exist a vector $|i' \ \textbf{\emph{b}}'\rangle \in  B_{x'}^{i'}$ such that $|M_{A}(i,\textbf{\emph{b}})^\top \rangle = |M_{A}(i',\textbf{\emph{b}}')^\top\rangle$. This is fulfilled by setting $(i', \textbf{\emph{b}}') = (i', \textbf{\emph{a}}') - (i, \textbf{\emph{a}}) + (i, \textbf{\emph{b}})$.

The second part follows immediately from the fact that we labeled the vectors in such a way that they are orthonormal to each other.
\end{proof}

\begin{lemma}
For every $i \in \mathbb{F}_{q}$ and $\textbf{\emph{x}} \in \im (M_{B})$, $|\phi_{x}^{i}\rangle$ is an eigenvector of $\rho_{A}$, which has norm equal to 1.
\end{lemma}
\begin{proof}
The density matrix for subsystem $A$ is given by
\begin{eqnarray}
\label{eq:rho_A}
\rho_{A} & = & \tr_{RB} |RP\rangle \langle RP| \nonumber \\
& = & \frac{1}{q^{e-1}} \sum_{i \in \mathbb{F}_{q}} \alpha_{i} \sum_{x \in \im (M_{B})} \nonumber \\
& & \qquad \sum_{|i \emph{\textbf{a}}\rangle, |i \emph{\textbf{a}}' \rangle \in B_{x}^{i}}
\Bigl| M_{A} \left( \begin{array}{@{}c@{}} i  \\ \emph{\textbf{a}} \end{array} \right) \Bigr\rangle \Bigl\langle M_{A} \left( \begin{array}{@{}c@{}} i \\ \emph{\textbf{a}}' \end{array} \right) \Bigr| \nonumber\\
& = & \frac{1}{q^{m}} \sum_{i \in \mathbb{F}_{q}} \alpha_{i} \sum_{x \in \im(M_{B})} |\phi_{x}^{i}\rangle\langle\phi_{x}^{i}|.
\end{eqnarray}
Since $M$ has independent columns and therefore its kernel only contains the all zero vector, the vectors $|\phi_{x}^{i}\rangle$ are correctly normalized. Because of Lemma \ref{lemma:oneelementincommon=>equalityB}, we have that the vectors $|\phi_{x}^{i}\rangle$ are all (not necessarily different) eigenvectors of $\rho_{A}$, which completes the proof. \end{proof}

In the next theorem, we compute the entropy of $\rho_{A}$ by calculating the eigenvalues of the eigenvectors of $\rho_{A}$.
\begin{lemma}
\label{theorem:entropysystemA}
Let the matrix $M$ have $e$ independent columns and let the rank of matrices $M_{A}$ and $M_{B}$ be $l$ and $m$ respectively. Then we have
\begin{eqnarray}
\label{eq:entropysystemA}
\textsf{\emph{S}}(A) & = & \textsf{\emph{S}}(S) + (m+l-e) \log q; \\
\label{eq:entropysystemB}
\textsf{S}(B) & = & (m+l-e) \log q.
\end{eqnarray}
\end{lemma}
\begin{proof} Consider any vector $|\phi_{x}^{i}\rangle$ for $i \in \mathbb{F}_{q}$ and $\textbf{\emph{x}} \in \im(M_{B})$. Because of Lemma \ref{lemma:oneelementincommon=>equalityB} and the fact that the kernel of $M$ only contains the all-zero vector, this vector is repeated $q^{e-l}$ times in Eq. (\ref{eq:rho_A}). Moreover, because of the properties of the MSP and the fact that $A$ is an authorized set, for all these $q^{e-l}$ vectors $|\phi_{x'}^{i'}\rangle$ we have that $i'= i$. Therefore, we can write for $\rho_{A}$
\begin{eqnarray}
\rho_{A} & = & \frac{q^{e-l}}{q^{m}} \sum_{i} \alpha_{i} \sum_{t} |\phi_{t}^{i}\rangle\langle\phi_{t}^{i}|,
\end{eqnarray}
where the vectors $|\phi_{t}^{i}\rangle$, with $1\le t \le q^{m+l-e}$ and $1 \le i \le q$, are all different. Moreover, the vectors $|\phi_{t}^{i}\rangle$ are all eigenvectors of $\rho_{A}$, each with eigenvalue $\alpha_{i} / q^{m+l-e}$. Hence, it follows that the entropy of system $A$ is given by Eq. (\ref{eq:entropysystemA}). The proof for the entropy of system $B$ is omitted here.\end{proof}

Finally, we have the following.
\begin{theorem}
For any adversary structure $\mathcal{A}$, there exists a QSS realizing $\mathcal{A}$.
\end{theorem}
\begin{proof}
We only prove the case that $\mathcal{A}$ is self-dual, the scheme for the other adversary structures can be obtained from this one.
In the case that $\mathcal{A}$ is self-dual, consider the scheme given by Eq. (\ref{eq:MSPconstruction}). Let $A \subseteq P, A \notin \mathcal{A}$ be an authorized set and $B = P \setminus A$. Then because of Lemma \ref{theorem:entropysystemA}, we have
\begin{eqnarray*}
\textsf{\emph{I}}(R:S) & = & \textsf{\emph{S}}(R) + \textsf{\emph{S}}(S) - \textsf{\emph{S}}(RS) =  2 \textsf{\emph{S}}(S);  \\
\textsf{\emph{I}}(R:A) & = & \textsf{\emph{S}}(R) + \textsf{\emph{S}}(A) - \textsf{\emph{S}}(B) =  2 \textsf{\emph{S}}(S),
\end{eqnarray*}
where we have used the fact that systems $RS$ and $RAB$ are in a pure state. The secrecy requirement is equivalent to the recoverability requirement in this case, but can also be checked directly.
\end{proof}

\section{QSS using Teleportation}
We verify the correctness of the $((n,n))$ QTS scheme using teleportation as was proposed in \cite{Bandyopadhyay00}. This is done by defining an equivalent scheme that does not use teleportation. \\

Let the state of the secret $S$ be given by the density matrix $\rho_{S} = \alpha_{0} |0\rangle \langle 0| + \alpha_{1} |1\rangle \langle 1|$, where $\alpha_{0}, \alpha_{1} \in \mathbb{C}$. The state of $S$ together with its reference system $R$ is then given by
\begin{equation}
|RS\rangle = \sqrt{\alpha_{0}} |00\rangle + \sqrt{\alpha_{1}} |11\rangle.
\end{equation}
Suppose the dealer $D$ and the $n$ players $P$ initially share the maximally entangled state
\begin{equation}
|\psi\rangle_{DP} = \frac{1}{\sqrt{2}} (|0\underbrace{0 \ldots 0}_{n}\rangle + |1\underbrace{1 \ldots 1}_{n}\rangle).
\end{equation}
The first step in the teleportation scheme is that the dealer lets the secret interact with his part of the entangled state and then performs a Bell measurement on his two qubits. If he then communicates the (classical) outcome of this measurement to the players, they are able to obtain the state
\begin{equation}
|RP\rangle = \sqrt{\alpha_{0}} |0\underbrace{0 \ldots 0}_{n}\rangle + \sqrt{\alpha_{1}} |1\underbrace{1 \ldots 1}_{n}\rangle.
\end{equation}

In \cite{Bandyopadhyay00} it was shown how the players can obtain the state of the secret if all of them cooperate. However, it was not analyzed what happens if a group of less than $n$ players cooperate. This is done here by formulating an equivalent protocol in terms of the information theoretical model. Let the isometry $V_{n,n} : \mathbb{C}^{2} \rightarrow (\mathbb{C}^{2})^{\otimes n}$  be defined by
\begin{equation}
V_{n,n} ( a|0\rangle + b |1\rangle) = a|\underbrace{0 \ldots 0}_{n}\rangle + b |\underbrace{1 \ldots 1}_{n}\rangle,
\end{equation}
where $a, b \in \mathbb{C}$. The encoding of the secret by using teleportation is then equivalent to applying the mapping $I_{R} \otimes V_{n,n}$ to the system $RS$, where $I_{R}$ is the identity mapping on system $R$. However, the difference is, that with this mapping the dealer actually has to send quantum shares to the players, while otherwise he only has to perform a Bell measurement and sending two classical bits.

Next, we calculate the mutual informations in order to determine which sets of players are authorized. Let $P_{i}$ be the system of player $i = 1, 2, \ldots, n$. Then
\begin{eqnarray}
\rho_{P_{1}} & = & \alpha_{0} |0\rangle \langle 0| + \alpha_{1} |1\rangle \langle 1| \\
\rho_{P_{12}} & = & \alpha_{0} |00\rangle \langle 00| + \alpha_{1} |11\rangle \langle 11| \\
& \vdots & \nonumber \\
\rho_{P_{12 ... n}} & = & \alpha_{0} |0 ...0\rangle \langle 0 ... 0| + \alpha_{1} |1 ... 1\rangle \langle 1 ... 1|,
\end{eqnarray}
hence the entropy of the system of an arbitrary set of players equals the entropy of the secret. For the mutual informations, we have
\begin{eqnarray}
\textsf{\emph{I}}(R:P_{1}) & = & \textsf{\emph{I}}(R:P_{12})  \ = \ \ldots =
\nonumber \\
\textsf{\emph{I}}(R:P_{12 ... n-1}) & = & \textsf{\emph{S}}(S) \nonumber \\ &<& \textsf{\emph{I}}(R:S);\\
\textsf{\emph{I}}(R:P_{12 ... n}) & = & 2 \textsf{\emph{S}}(S) \nonumber \\ & = & \textsf{\emph{I}}(R:S),
\end{eqnarray}
since $RP_{12 ... n}$ is the only system with entropy not equal to $\textsf{\emph{S}}(S)$, but equal to 0.
Hence, a set of less than $n$ players has some information about the secret, but not enough to recover it, while all $n$ players together have enough information to recover the secret. Therefore, we have shown that this scheme is a non-perfect $((n,n))$ QTS.

\end{document}